\documentclass[12pt,preprint]{emulateapj}

\submitted{Accepted for the Astrophysical Journal}

\begin{document}

\title{A deep Chandra observation of Abell 4059: a new face to
``radio-mode'' AGN feedback?} \author{Christopher S.
Reynolds\altaffilmark{1}, Elyse A. Casper,\altaffilmark{1,2} and
Sebastian Heinz\altaffilmark{3}} \altaffiltext{1}{Astronomy
Department, University of Maryland, College Park, MD 20742}
\altaffiltext{4}{Think Energy Inc., 6930 Carroll Avenue, Takoma Park, MD 20912}
\altaffiltext{3}{Astronomy Department, University of Wisconsin,
Madison, WI 53706}

\begin{abstract} 
A deep Chandra observation of the cooling core cluster Abell 4059
(A4059) is presented.  Previous studies have found two X-ray cavities
in the central regions of A4059 together with a ridge of X-ray
emission 20\,kpc south-west of the cluster center.  These features are
clearly related to the radio galaxy PKS~2354--35 which resides in the
cD galaxy.  Our new data confirm these previous findings and
strengthen previous suggestions that the south-western ridge is colder
and denser than, but in approximate pressure equilibrium with, the
surrounding ICM atmosphere.  In addition, we find evidence for a weak
shock that wraps around the north and east sides of the cavity
structure. Our data allow us to map the 2-dimensional distribution of
metals in the ICM of A~4059 for the first time.  We find that the SW
ridge possesses an anomalously high (super-solar) metalicity.  The
unusual morphology, temperature structure and metal distribution all
point to significant asymmetry in the ICM atmosphere prior to the
onset of radio-galaxy activity.  Motivated by the very high
metalicity of the SW ridge, we hypothesize that the ICM asymmetry was
caused by the extremely rapid stripping of metal enriched gas from a
starburst galaxy that plunged through the core of A4059.  Furthermore,
we suggest that the onset of powerful radio-galaxy activity in the cD
galaxy may have been initiated by this starburst/stripping event,
either via the tidal-shocking of cold gas native to the cD galaxy, or
the accretion of cold gas that had been stripped from the starburst
galaxy.
\end{abstract}


\section{Introduction}
\label{introduction}

In the core regions of galaxy clusters, X-ray observations directly
reveal the radiative cooling of the intracluster medium (ICM).  In
most relaxed clusters, the radiative cooling time is measured to be
substantially less than the Hubble time (see \citet{peterson06}). In
the absence of a balancing heat source, this cooling would result in a
slow flow of matter inwards towards the center of the cluster, a
cooling flow \citep{cowie77,fabian77,fabian94}, and a growing central
mass of cooled gas.  The absence of signatures of this cold gas at
non-X-ray wavelengths leads to the cooling flow problem.  More recently,
high-resolution X-ray spectroscopy with {\it XMM-Newton} has shown an
almost complete lack of gas below a temperature of $T_{\rm vir}/3$,
where $T_{\rm vir}$ is the virial temperature of the core regions of
the cluster \citep{peterson01,tamura01}.  It is possible that gas
below $T_{\rm vir}/3\sim 1-2\ {\rm keV}$ is present but not
detectable.  For example, Fabian et al. (2001) discusses a model in
which there are strong metalicity inhomogeneities and the high
metalicity gas cools below 1--2 keV too rapidly for detection. No
compelling model for the creation of such inhomogeneities exists,
however. Additionally, a genuine lack of cool gas would help explain
(and indeed may be {\it required} to explain) the deficiency of
massive galaxies that is revealed while comparing the galaxy mass
function and the dark matter halo mass function \citep{benson03}.

The cooling flow problem can clearly be solved if some phenomenon
heats the ICM core enough to, on average, balance radiative cooling.
A leading possibility is that jets from active galactic nuclei (AGN)
at the center of galaxy clusters somehow provide this heating. It has
been well established that AGN undergo complex interactions with the
surrounding ICM, resulting in cavities
\citep{fabian00,mcnamara00,blanton01,young02,nulsen05b}, ghost
cavities \citep{fabian00,mcnamara01,heinz02,choi04}, ripples
\citep{fabian03,fabian05}, shells \citep{fabian00}, shocks
\citep{jones04,nulsen05a,nulsen05b}, and filaments
\citep{nulsen05b}. It is feasible that one or several of these
interactions result in the required heating.

In this paper, we present a new and moderately deep Chandra
observation of the cooling core cluster Abell 4059 (A4059;
$z=0.049$). Previous X-ray studies have shown that A4059 has a double
peaked surface brightness distribution, with a bright ridge of X-ray
emission located to the southwest of the cluster center
\citep{huang98,heinz02,choi04}.  This ridge seems to be in pressure balance
with the ambient material.  At the western boundary of this ridge
there is a sharp gradient in temperature and density.  It was
suggested that this gradient could be explained by radiative cooling
resulting from an interaction between the ICM and the radio-galaxy
induced disturbance. While the metalicity profile of A4059 was not
clear from these previous studies, there was weak evidence that it
peaks at roughly solar values in the center regions and decreases
outwards \citep{choi04}.

Additionally, previous studies have shown two X-ray cavities in the
center of A4059. The northwest cavity is clearly visible and distinct,
while the southeast cavity is more difficult to discern
\citep{huang98,heinz02}. These previous studies found no evidence of
shocks around the cavities.  It was noted however that the axis
connecting the centers of the two cavities does not pass through the
position of the AGN itself, instead passing to the northeast of the
galactic nucleus by $\sim 10^{\prime\prime}$.  This prompted
\citet{heinz02} to suggest that the radio-galaxy outburst erupted into
an ICM that possessed a pre-existing bulk flow, and that the cavities
have been pushed in the NE direction by that flow.

High-frequency radio studies ($1.4-8.5$\,GHz) of this cluster indicate
the presence of two radio lobes \citep{choi04,taylor94} associated
with the radio galaxy PKS2354--35. The northwest lobe only partially
covers the northwest X-ray cavity, and the southeast lobe simply
misses the southeast X-ray cavity. Both X-ray cavities are therefore
ghost cavities, probably formed from a previous powerful burst of
activity by the central FR I radio galaxy, PKS 2354-35
\citep{heinz02,choi04}.  This has been recently confirmed by the
detection of low frequency (330\,MHz) radio emission coincident with
these cavities (T.~Clarke, private communication).

As we discuss in this paper, our new {\it Chandra} data change our
view of this cluster.  Section~\ref{obs} describes the {\tt \it
Chandra} observations and basic data reduction. Section~\ref{results}
discusses additional data reduction and presents X-ray images,
temperature and metalicity maps, and the results of the deprojection
analysis. The results are discussed and interpreted in
\S~\ref{discussion}.  In particular, we discuss how considerations of
the metallicity asymmetry in A4059 leads us to a rather novel view of
radio-mode feedback in this cluster.  Finally, we present our
conclusions in Section~5.  Throughout this paper, we use J2000
coordinates and assume a standard WMAP cosmology (flat universe with
$\Omega_{\Lambda}=0.73$, $H_0=71\,{\rm km}\,{\rm s}^{-1}\,{\rm
Mpc}^{-1}$; Spergel et al. 2003). Given the redshift of
A4059/PKS2354--35 of $z=0.04905$, this gives a luminosity distance of
$215.1$\,Mpc and a scale of 0.95\,kpc per arcsecond
\citep{wright06}\footnote{http://www.astro.ucla.edu/$\sim$wright/CosmoCalc.html}.

\section{Observations and Data Reduction} 
\label{obs}

{\it Chandra} observed Abell~4059 on three separate occasions;
24-September-2000 (total exposure 24.6\,ksec), 4-January-2001 (total
exposure 20.1\,ksec), and 26/27-January-2005 (total exposure
93.6\,ksec). Results from the first two, shorter, observations have
already been presented by \citet{heinz02} and \citet{choi04}. All
observations were performed with the Advanced CCD Imaging Spectrometer
(ACIS) placing the core of the cluster close to the telescope
aim-point on the central back-illuminated chip (i.e., chip S3). All
data were reduced with CIAO version-3.4.  The level-1 data were
processed in order to apply the latest gain correction map as well as
the latest version of the charge transfer inefficiency (CTI)
correction.  The processed level-1 data were then filtered with a
standard grade selection (including ASCA grades 0,2,3,4,6) resulting
in new level-2 events lists. Periods of high background were searched
for by examining the chip-S1 lightcurve. The September-2000 data were
found to be affected by a large background flare; elimination of the
affected data reduced the good exposure time to 16\,ksec. The 2001 and
2005 data were not affected by significant background flares, allowing
the full 24.6\,ksec and 93.6\,ksec, respectively, to be utilized.

The three datasets (totaling 134.2\,ksec of usable data) were merged
using the {\tt merge\_all} script in CIAO.  A 4.2' {$\times$} 4.2'
(238 kpc {$\times$} 238 kpc) region enclosing the cluster core was
extracted to create an exposure-corrected image.  It would, however,
be inappropriate to use the merged dataset for spectral studies. The
spectral response of the ACIS depends on the chip coordinates, which
map to different physical coordinates for the three
observations. Additionally, the spectral response has degraded with
time due to a slow increase in charge transfer inefficiency (CTI) and
the build-up of a contaminant on the ACIS filter. For these reasons,
spectral studies are only performed on the 2005 data.  Given that the
2005 observation corresponds to 70\% of our total good data for this
object, this restriction does not significantly impact the
signal-to-noise of our spectral study.

\section{Analysis and Results}
\label{results}

\subsection{X-ray Images}
\label{images}

\begin{figure*}
\centerline{
\includegraphics[width=1.0\textwidth]{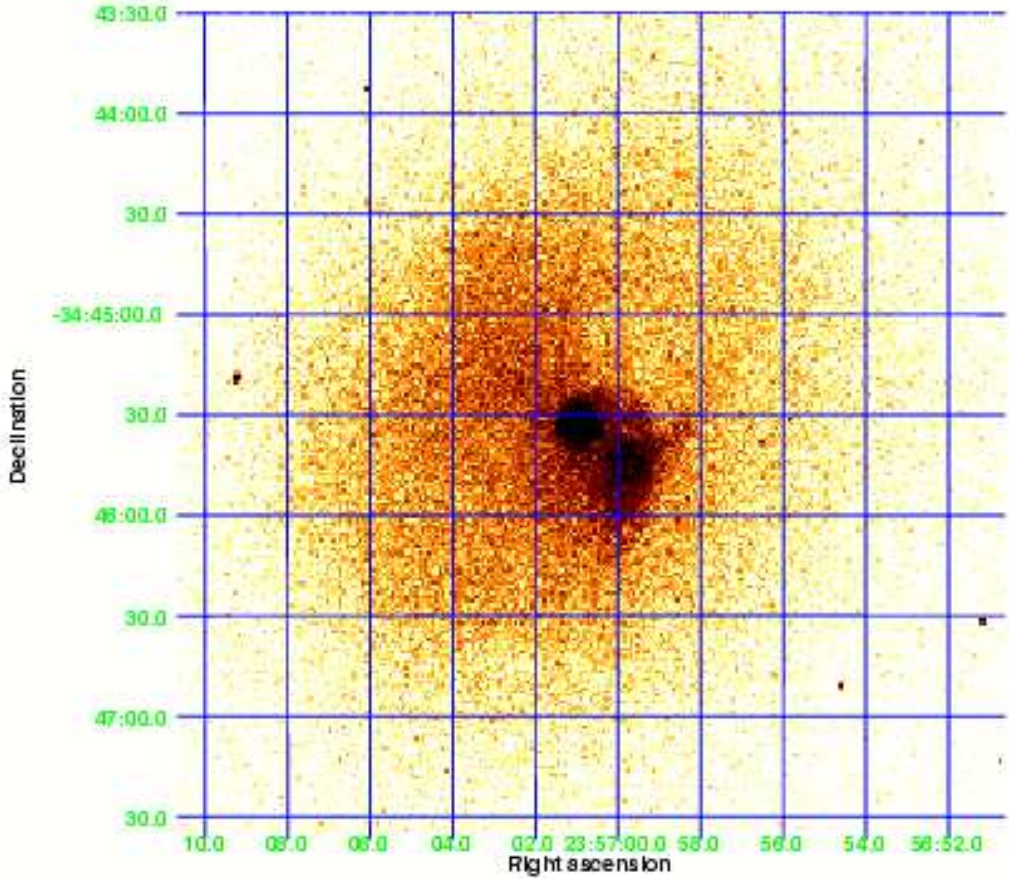}
}
\caption{Exposure-corrected full-band (0.3--10\,keV) image of A4059.}
\label{fig:raw}
\end{figure*}

Figure~\ref{fig:raw} shows the exposure-corrected full-band
(0.3--10\,keV) unsmoothed image of A4059.  The double peak
(corresponding to the cluster core and the SW ridge) is prominent, and
the northwest cavity is clearly visible. A spur of bright emission is
seen to the southwest of the core.  The improved signal-to-noise allow
previously unknown aspects of the morphology to be discerned.  The
core of the cluster and the SW ridge are clearly connected in the
north by a high-contrast curved rim.  This rim appears to continue
through the brightest part of the SW ridge and define the southern
parts of the ridge.  Although much less distinct, there may be a
similar structure which is the approximate reflection of this rim in
the line joining the cluster core and the brightest part of the SW
ridge.  At a much lower contrast, one can trace a thin plume extending
from the core directly northwards for 60\,kpc (1\,arcmin), and another
plume extending northwest from the SW ridge.

\begin{figure*}
\centerline{
\includegraphics[width=1.0\textwidth]{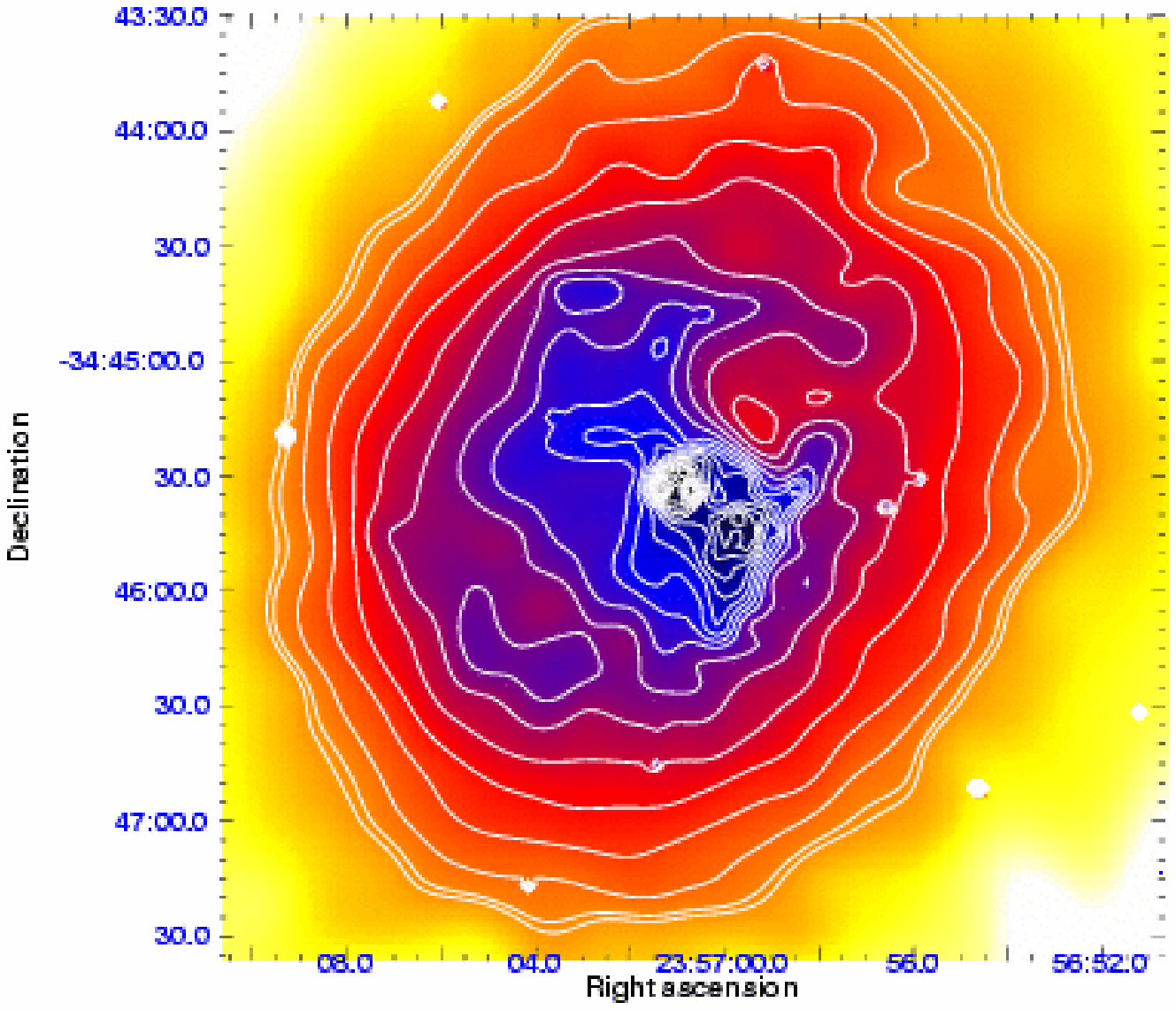}
}
\caption{Adaptively-smoothed, exposure corrected image of the core
regions of A4059.  Logarithmically-spaced contours of this same map
are also over-laid.}
\label{fig:adapt}
\end{figure*}

From the merged data, we produced adaptively smoothed,
exposure-corrected, full band (0.3-10 keV) images using the CIAO tool
{\tt csmooth}.  Significance levels in the range $3-5\sigma$ were used
to set the smoothing scales.  The adaptively-smoothed image shown in
Fig.~\ref{fig:adapt} supports earlier morphological studies of Abell
4059. While the outer portion of the cluster is fairly relaxed, the
inner 60 kpc is rather disturbed showing the cavities and south-west
ridge that have been well studied in previous works.  Our deeper image
reveals new structure, however.  The adaptively smoothed image hints
at part of a shell bounding the disturbed region of the cluster,
especially to the southeast of the cluster.

\begin{figure*}
\centerline{
\includegraphics[width=0.6\textwidth]{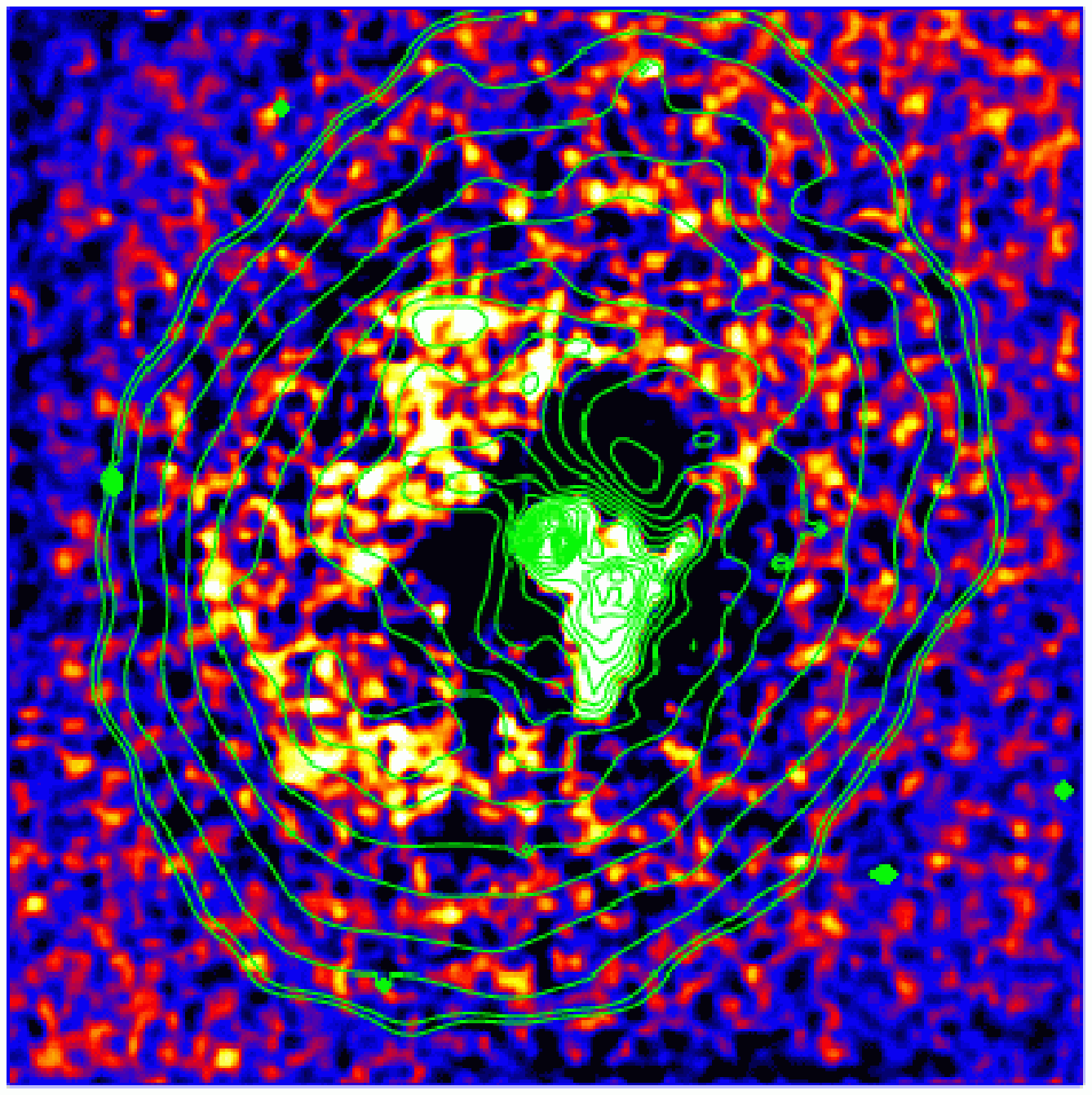}
}
\caption{Unsharp mask image of A4059, created by subtracting a full
band map smoothed with a $\sigma=12.5^{\prime\prime}$ Gaussian from a
map smoothed with a $\sigma=1.5^{\prime\prime}$ Gaussian.  Overlaid
are contours of the adaptively-smoothed map.}
\label{fig:unsharp}
\end{figure*}

To investigate this apparent shell in more detail, we created an
unsharp-mask image following the approach of Fabian et al. (2003).  In
detail, images with two different Gaussian smoothings were created
from the merged data, one image with a Gaussian smoothing scale of 25
pixels ($12.5^{\prime\prime}$) and the other with a Gaussian smoothing
scale of 3 pixels ($1.5^{\prime\prime}$). The unsharp-mask image was
created by subtracting the heavily smoothed image from the lightly
smoothed image, a process which allows one to search for subtle
discontinuities in either the surface brightness or the gradient of
the surface brightness.  The result is shown in
Fig.~\ref{fig:unsharp}, and reveals a subtle ``edge'' which has an
approximate semi-circular form in the eastern and northern quadrants.
The projected radius of this feature is 70--80\,kpc.  There are hints
that, once this edge crosses the jet axis, this feature may sweep
inwards and connect onto each end of the SW ridge.  

\begin{figure}[b]
\vspace{1cm}
\centerline{
\includegraphics[width=0.35\textwidth,angle=270]{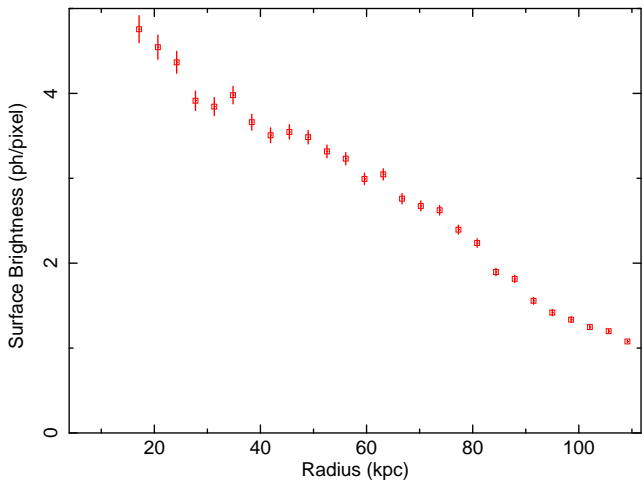}
}
\caption{Surface brightness profile for a sector of the cluster
between position angles 225$^\circ$ and 270$^\circ$ (i.e., the ESE
sector).}
\vspace{0.5cm}
\label{fig:radial_profile}
\end{figure}

Another view of this structure is obtained by constructing radial
surface brightness profiles.  Figure~\ref{fig:radial_profile} shows
the surface brightness profile for a sector of the cluster between
position angles $225^\circ$ and $270^\circ$.  This sector cuts the
edge noted in the unsharp-mask image at a distance of 80\,kpc from the
cluster center and, indeed, a small discontinuity in gradient of the
surface brightness can be seen at $r=80\,{\rm kpc}$.  Unfortunately,
further studies of this feature (e.g., searching for temperature jumps
associated with the surface brightness jump) are beyond what is
possible with the current dataset.  Rigorously characterizing the
nature of this feature must await sigificantly deeper X-ray data.
However, by analogy with findings in other systems (e.g., Virgo;
Forman et al. 2007), it seems likely that this feature is a weak shock
or compression wave driven into the ICM by the radio galaxy activity.
In \S~\ref{discussion}, we discuss the implications of this hypothesis
for the age of the AGN.  As noted in Markevitch et al. (2000), a
curved shock/compression front viewed in projection does {\it not}
produce a true discontinuity in the surface brightness but, instead,
is revealed through a discontinuity in the gradient of the surface
brightness of the kind suggested by Fig.~\ref{fig:radial_profile}.

\subsection{Imaging Spectroscopy Using Adaptive Binning}

By examining the spectrum of the X-ray emission across the image, the
spatial distribution of temperature, metalicity, density, pressure,
and entropy can be probed.  This kind of physical information is
crucial if we are to disentangle this complex source.  In particular,
understanding the spatial distribution of metals in this cluster is a
major motivation for these new data.  

Spectral analysis was performed on data that had been adaptively
binned using the Weighted Voronoi Tesselation (WVT) binning algorithm
of \citet{diehl06}\footnote{http://www.phy.ohiou.edu/$\sim$ diehl/WVT}. This
algorithm produces polygonal bins with a constant signal to noise
ratio (SNR), where the number of photons in each bin is the square of
the prescribed SNR.  Once the optimal polygon-tiling of the image
plane has been determined, the CIAO tool {\tt specextract} was used to
extract spectra, energy response matrices and effective area curves
for each tile.  A single background spectrum was obtained from the
source free region at the north-east corner of the S3-chip.  After
experimentation, we determined that a SNR of 30 allows us to construct
adequate maps of plasma temperature and emission measure (and hence
density, pressure and entropy).  Metalicity maps requires higher
quality spectra and could only be constructed once the SNR was
increased to 40.

The spectrum from each bin was modeled using a single temperature
thermal plasma model {\tt mekal}
\citep{mewe85,mewe86,kaastra92,liedahl95} modified by the effects of
Galactic photoelectric absorption with column density $N_{\rm
H}=1.45\times 10^{20}\,{\rm cm}^{-2}$.  The abundance of metals within
the plasma was a free parameter of the spectral models.  However, the
relative abundance of all elements heavier than helium was fixed at
their cosmic ratios (Anders \& Grevesse 1989); no statistical
improvement in the fit is obtained by relaxing this assumption.
Spectral fitting was performing using XSPEC version 11.3.2.  Although
we formally fit the 0.5--10\,keV data, the spectra of most tiles
become background dominated about 7\,keV.  The absorption column was
held fixed for the spectral fitting presented here, although we have
confirmed that our results are robust to allowing $N_{\rm H}$ to be a
free parameter.

\begin{figure*}
\centerline{
\hbox{
\includegraphics[width=0.5\textwidth]{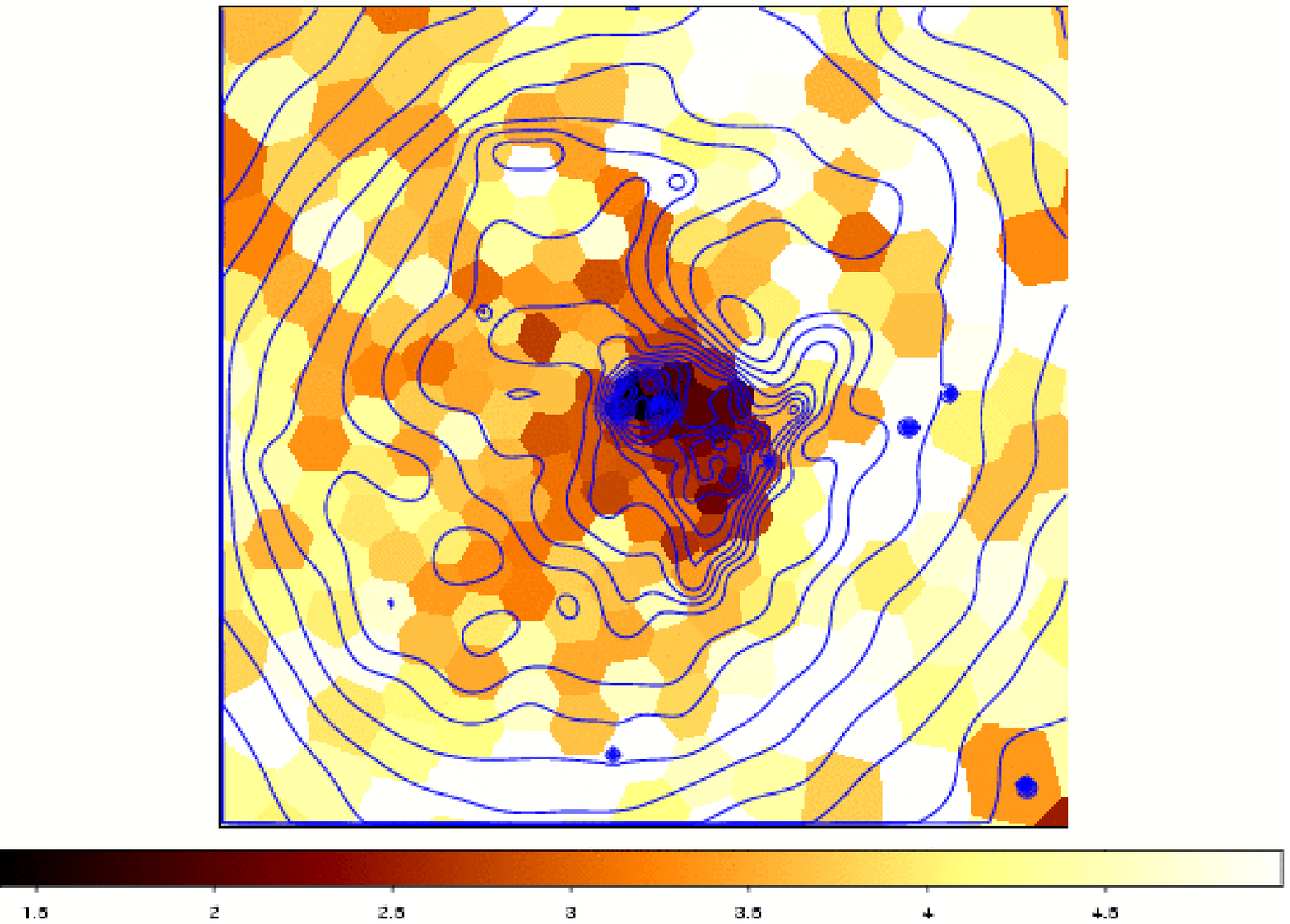}
\includegraphics[width=0.5\textwidth]{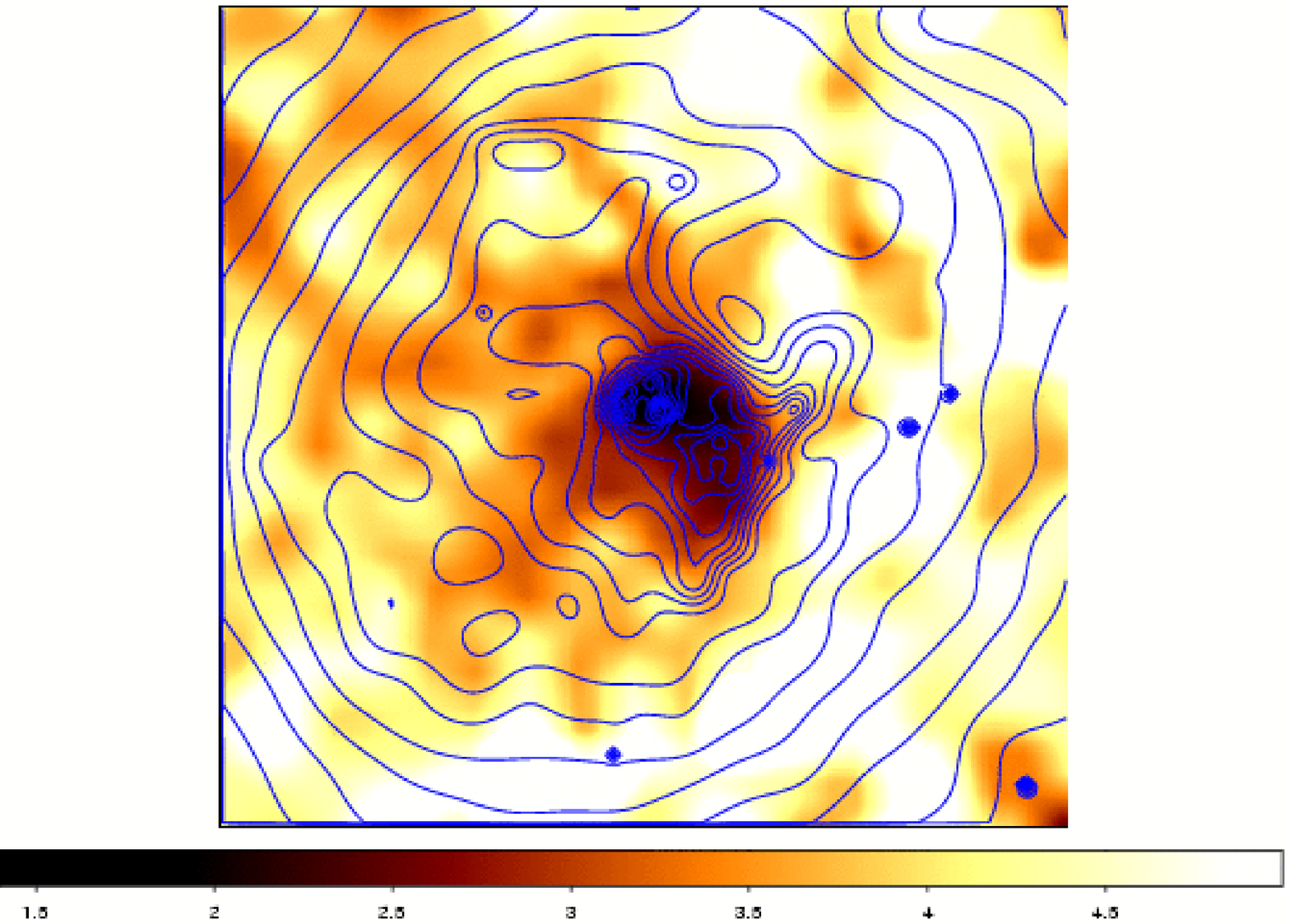}
}
}
\centerline{
\hbox{
\includegraphics[width=0.5\textwidth]{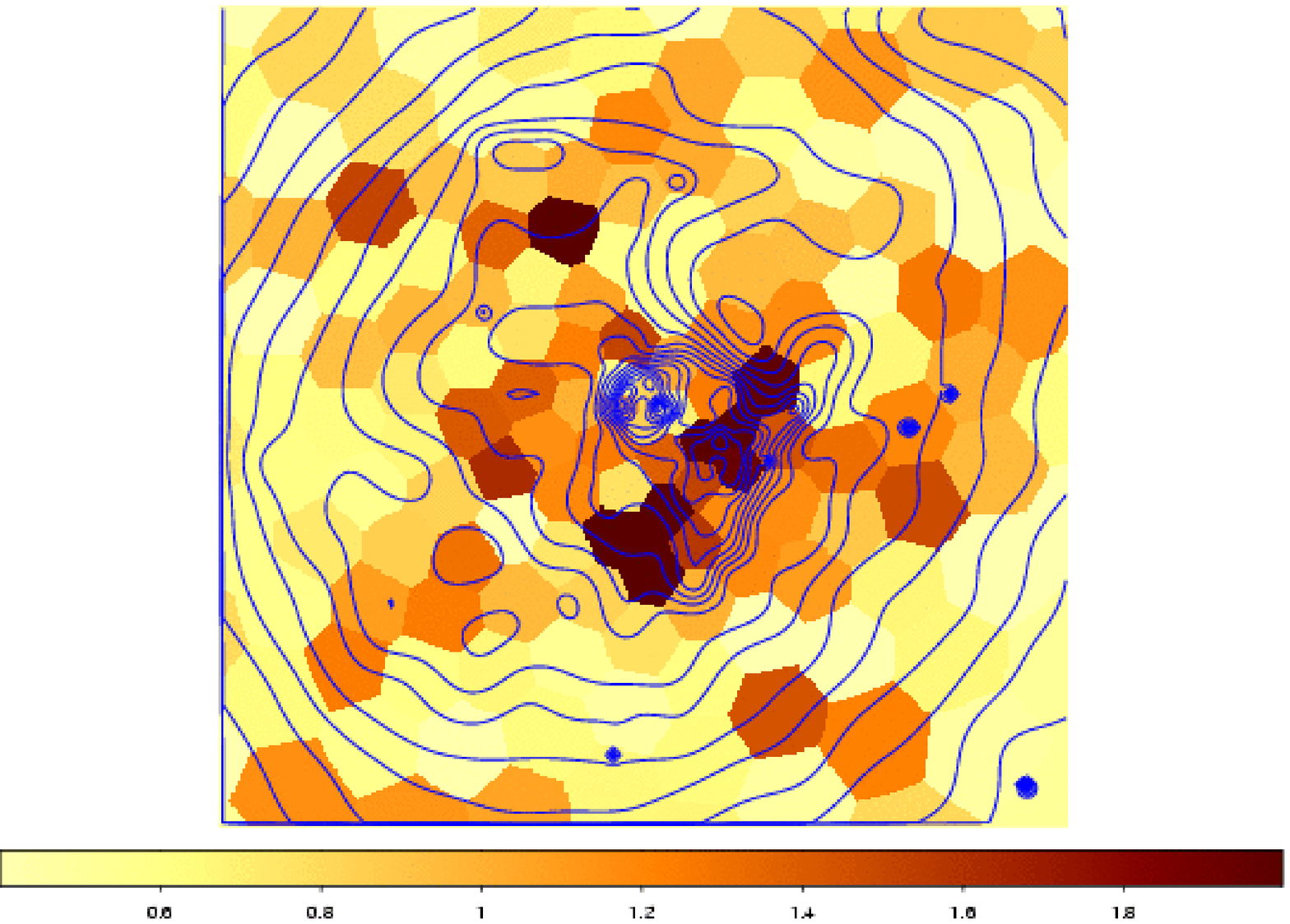}
\includegraphics[width=0.5\textwidth]{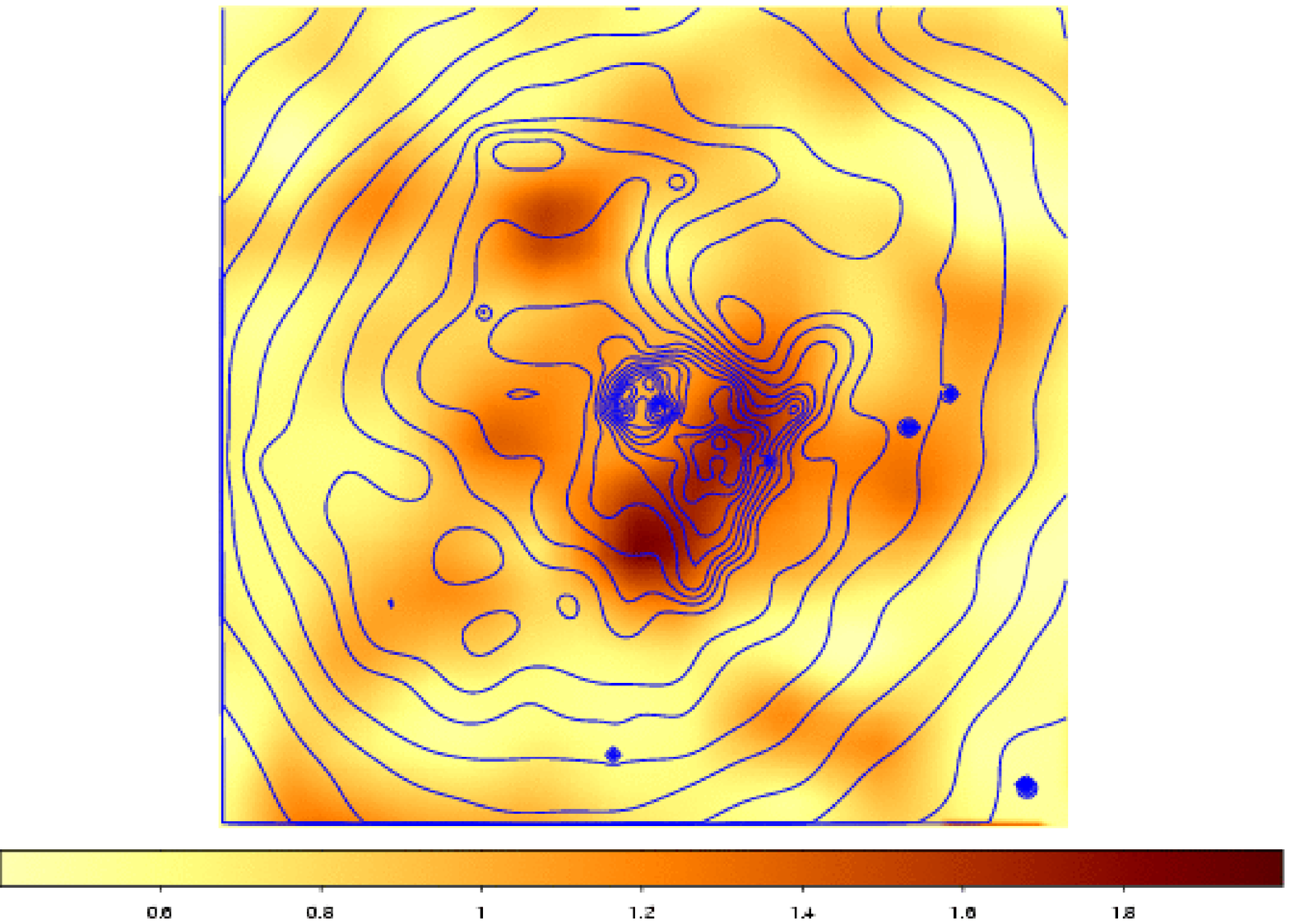}
}
}
\centerline{
\hbox{
\includegraphics[width=0.5\textwidth]{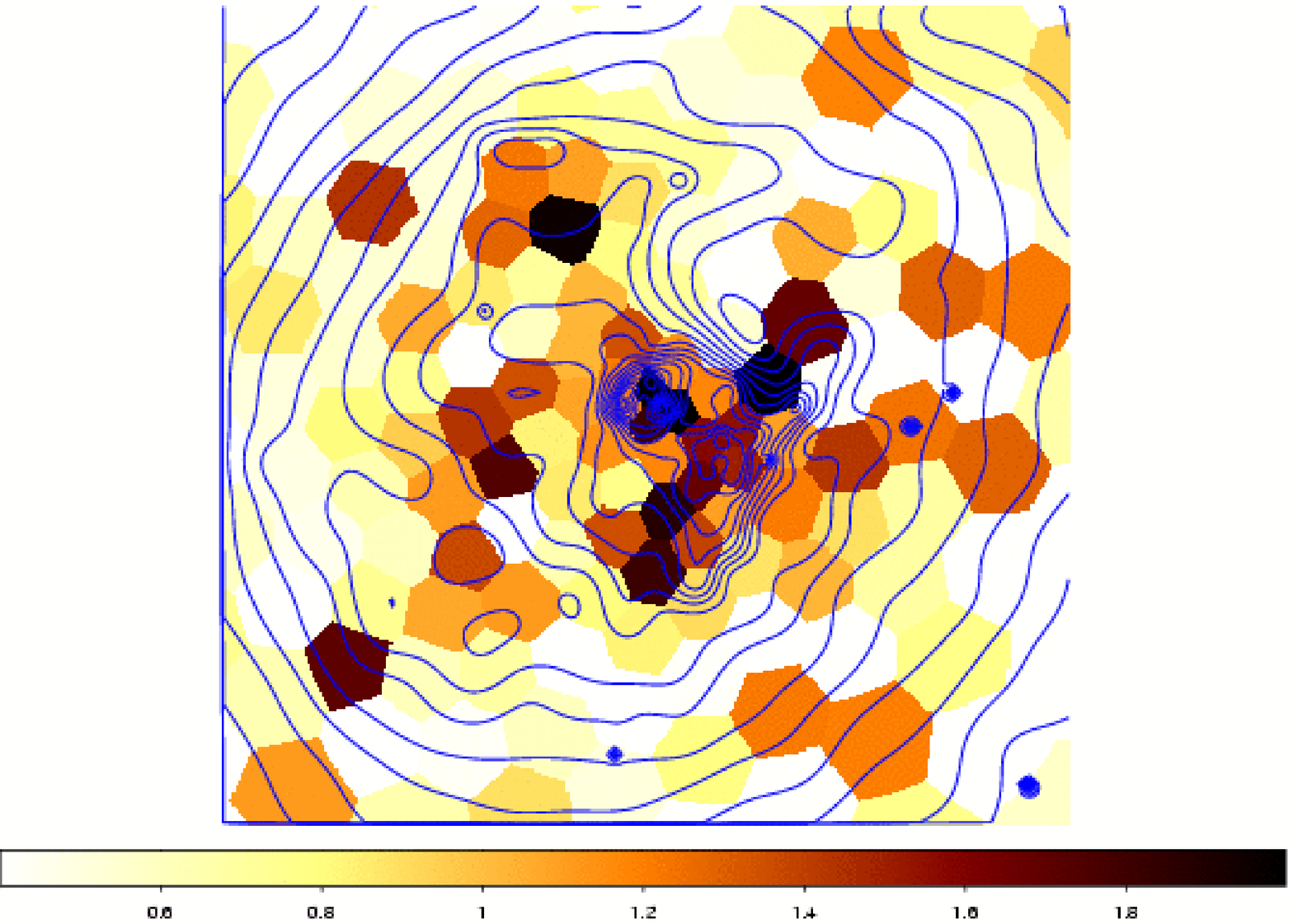}
\includegraphics[width=0.5\textwidth]{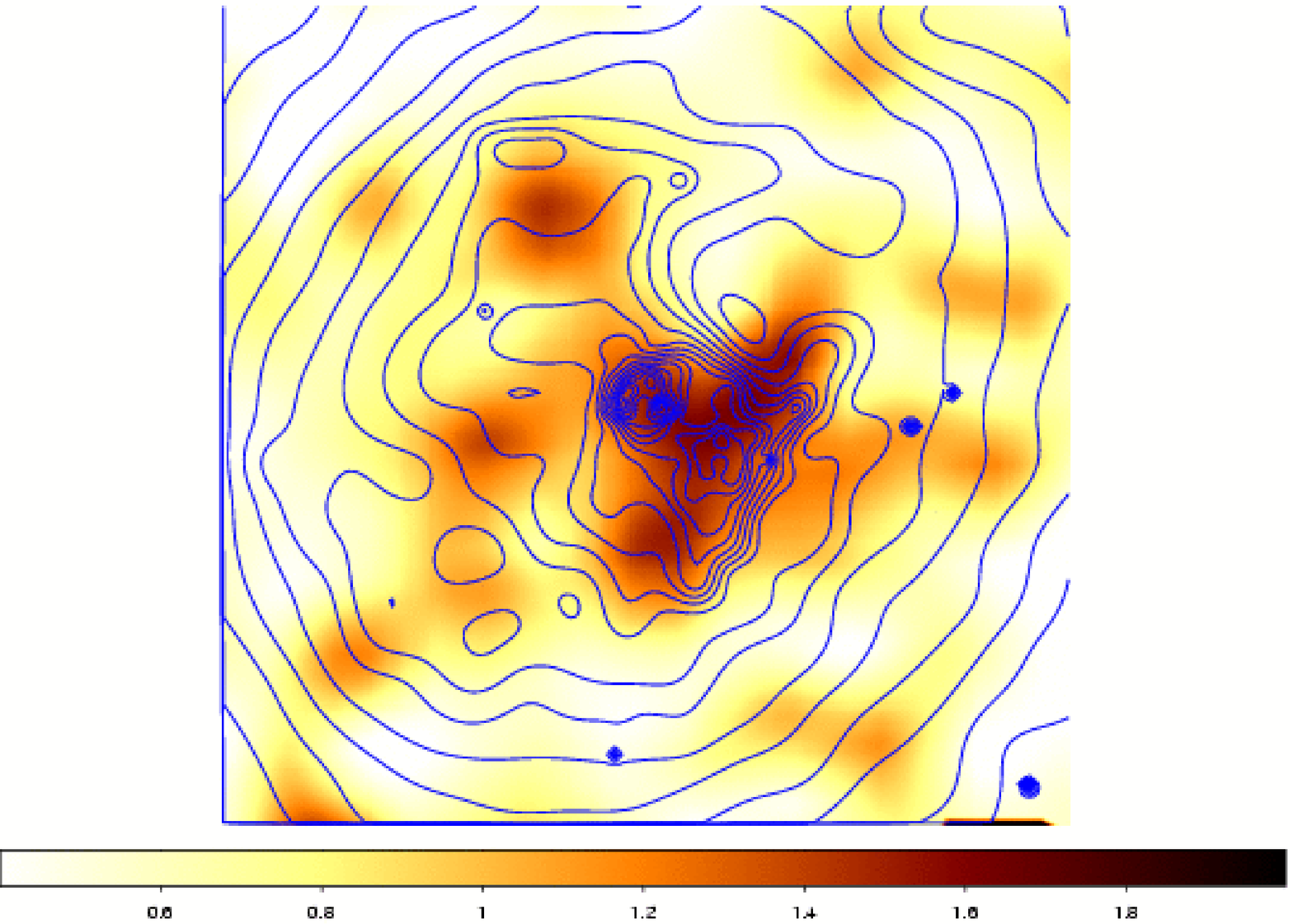}
}
}
\caption{{\small {\it Top left panel : }Temperature map of the core regions of
A4059 based on the ${\rm SNR}=30$ Voronoi Tesselation.  Color-bar
units are keV.  {\it Top right panel : }Smoothed temperature map,
using a $20^{\prime\prime}$ box-car smoothing kernel.  {\it Middle
left panel : }Metalicity map (based on single temperature fits) of the
core regions of A4059 based on the ${\rm SNR}=40$ Voronoi Tesselation.
Color-bar units are cosmic abundances (as defined by Anders \&
Grevesse 1989).  {\it Middle right panel : }Smoothed (single
temperature) metalicity map, using a $40^{\prime\prime}$ box-car
smoothing kernel. {\it Bottom left panel : }Metalicity map (based on
two-temperature fits) of the core regions of A4059 based on the ${\rm
SNR}=40$ Voronoi Tesselation.  Color-bar units are cosmic abundances
(as defined by Anders \& Grevesse 1989).  {\it Bottom right panel :
}Smoothed (two-temperature) metalicity map, using a
$40^{\prime\prime}$ box-car smoothing kernel. }}
\label{fig:temp_metal}
\end{figure*}

The temperature map (made from the ${\rm SNR}=30$ tesselation) is
shown in Fig.~\ref{fig:temp_metal} (top panels).  The most obvious
feature in this map is the region of cool ($<2.5\,{\rm keV}$) gas that
extends from the core through the SW ridge.  Examination of this
temperature map shows that cool gas has a close spatial correspondence
with the highest surface brightness regions of the SW ridge.  In
particular, the boundary of the region containing the cool gas closely
follows the high-contrast curved rim of emission that connects the
core with the SW ridge (see discussion in \S~\ref{images}), and
continues to track this curved rim through the SW ridge and into the
southern extremity of the SW ridge.  Interestingly, the spur that
projects from the SW ridge in the north-westerly direction is
appreciably hotter than the rest of the SW ridge (3.5--4.0\,keV as
opposed to 2.0--2.5\,keV).

For the first time, these data allow us to examine the 2-dimensional
metalicity distribution for this cluster.  The metalicity map derived
from single temperature thermal plasma fits (using the ${\rm SNR}=40$
map) is shown in Fig.~\ref{fig:temp_metal} (middle panels).  The most
significant feature in this map is the dramatic enhancement in
metalicity (upto twice the average cosmic abundance) along the SW
ridge. Metal enriched gas appears to be present along the entire
length of the SW ridge, including the north-westerly spur which was
noted above to be at a significantly higher temperature than the rest
of the SW ridge.  It is particularly striking in this map that the
metalicity of the SW ridge appears to be appreciably higher than that
of the central core itself.

The picture concerning the metalicity distribution in the centralmost
regions changes somewhat, however, when we go beyond a single
temperature thermal plasma model for the spectrum of each bin.  In
particular, we have fitted spectra from the tiles of the ${\rm
SNR}=40$ tesselation with a two-component thermal plasma model; the
temperature and emission measure of each plasma component is a free
parameter of the fit, but the two components are assumed to share a
common metalicity.  Comparing the $\chi^2$ goodness of fit parameter
for the single-component and two-component plasma models, we find that
the two-component model is not a statistically better description of
the data than the single-component model (employing the F-test for two
additional model parameters) for the spectrum from {\it most} of the
tiles, {\it including those that make up the SW ridge}.  Furthermore,
for this vast majority of tiles, the best-fitting value of the
metalicity parameter is similar between the single-temperature and
two-temperature fits.  This last result is unsurprising since, in most
cases, the normalization (emission measure) of one of the plasma
components in these two-component fits vanishes, leaving essentially a
single temperature fit.  The spectra from the tiles in the centralmost
bins of the cluster are, however, described significantly better (at
the 90\% or more level) by the two-component plasma model.  The need
for these multi-temperature models may reflect the morphological
complexity of the ICM center or the combined contributions from both the
core of the ICM and the ISM of the cD galaxy itself.  

It is important to assess the effect of the multi-component model on
the inferred temperature and metalicity distributions.  We find that
the one-component temperature maps shown in Fig.~\ref{fig:temp_metal}
(top panels) remain an accurate approximation to the {\it
emission-measure weighted} temperature map derived from the
two-component fit.  On the other hand, the metalicity of the core
inferred from the two-component fits is appreciably higher than that
inferred from the single-temperature fits ($Z=1.2-2Z_\odot$ compared
with $Z=0.6-1.0Z_\odot$).  The metalicity map derived from the
two-component plasma fits Fig.~\ref{fig:temp_metal} (bottom panels)
still show a notable enhancement of metalicity along the SW ridge, but
the metalicity enhancement now also extends into the regions
immediately surrounding the cD galaxy.

\begin{figure}
\centerline{
\includegraphics[width=0.35\textwidth,angle=270]{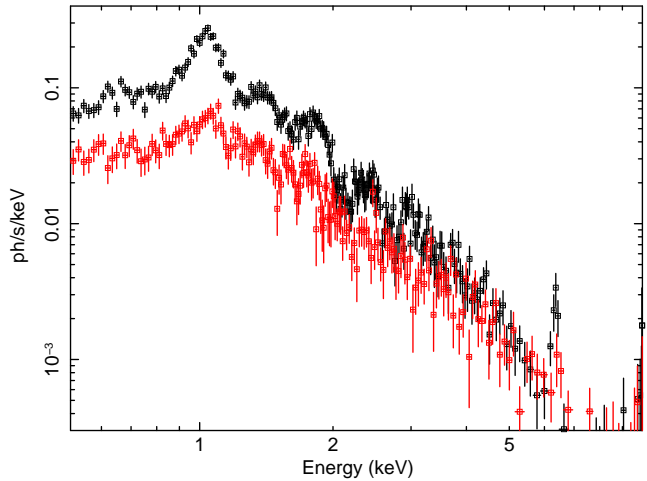}
}
\caption{The 0.5--10\,keV spectra of an elliptical region encompassing
the SW ridge (upper spectrum; black points) and an off-ridge region of
the same size located $12^{\prime\prime}$ to the south-west of the SW
ridge (lower spectrum; red points).  Note that the iron-L complex at
1\,keV and the iron-K line at $\sim 6$\,keV are appreciable stronger
in the ridge spectrum.}
\vspace{0.5cm}
\label{fig:2spec}
\end{figure}

To demonstrate the robustness of the conclusion regarding the high
abundance of the SW ridge, we have extracted the 0.5--10\,keV spectrum
from a single elliptical region ($25^{\prime\prime}\times
10^{\prime\prime}$) that encompasses the whole SW ridge and compared
it with the spectrum from the same sized region located
$12^{\prime\prime}$ to the south-west (i.e., beyond the SW ridge).
One and two temperature thermal plasma fits to the SW-ridge spectrum
implies a metallicity of $Z=1.70\pm 0.18Z_\odot$ and
$Z=1.84^{+0.29}_{-0.27}Z_\odot$, respectively.  On the other hand, one
and two temperature fits to the non-ridge spectrum yields $Z=0.87\pm
0.24Z_\odot$ and $Z=0.83^{+0.34}_{-0.26}Z_\odot$, respectively.
Indeed, this metallicity difference is visually apparent in the
spectra; examination of the spectra (Fig.~\ref{fig:2spec}) reveals an
iron-L line complex at $\sim $1\,keV and iron-K line at $\sim $6\,keV
which are clearly stronger in the ridge spectrum than the off-ridge
spectrum.

With some assumptions about the line-of-sight geometry, the emission
measure can be used to compute the characteristic density of the
plasma contributing to the observed radiation from a given spatial
bin.  Operationally, the electron number density $n_e$ is obtained
from the normalization of the {\tt mekal} component ($N$) in the
one-temperature fits by
   \begin{equation}
   n_e=D_{A}(1+z)\sqrt{4 \times 10^{14} \pi N V^{-1}} \label{eq:density},
   \end{equation}
where $D_{A}=195$\,Mpc is the angular size distance to A4059,
$z=0.049$ is the redshift, and $V$ is the effective volume of the
emitting plasma in that bin.  The volume $V$ was estimated by
multiplying the area of the Voronoi tile by the distance to the
cluster center, as defined by the location of peak brightness (which
coincides with PKS~2354--35 itself).  Given the density and spectrally
determined temperature, the pressure is obtained from the ideal gas
equation:
   \begin{equation}
   P= n k T \label{eq:pressure}.
   \end{equation}
where $n$ is the total particle number density of the plasma, assumed
to be given by $n=1.93n_e$ (appropriate for 75\% H and 25\% He).  The
entropy index was calculated using
   \begin{equation}
   s=\frac{k_BT}{n_e^{\gamma - 1}}. \label{eq:entropy}
   \end{equation}
where we assume a standard adiabatic index of $\gamma=5/3$.  Finally,
we compute the radiative cooling timescale
   \begin{equation}
   \tau_{\rm cool}\equiv\frac{3nkT}{2n_en_H\Lambda(T)}
   \end{equation}
where $n_H$ is the hydrogen number density, and $\Lambda(T)$ is the
cooling function modeled using the fitting formula of Sutherland \&
Dopita (1993).

\begin{figure*}
\hbox{
\includegraphics[width=0.5\textwidth]{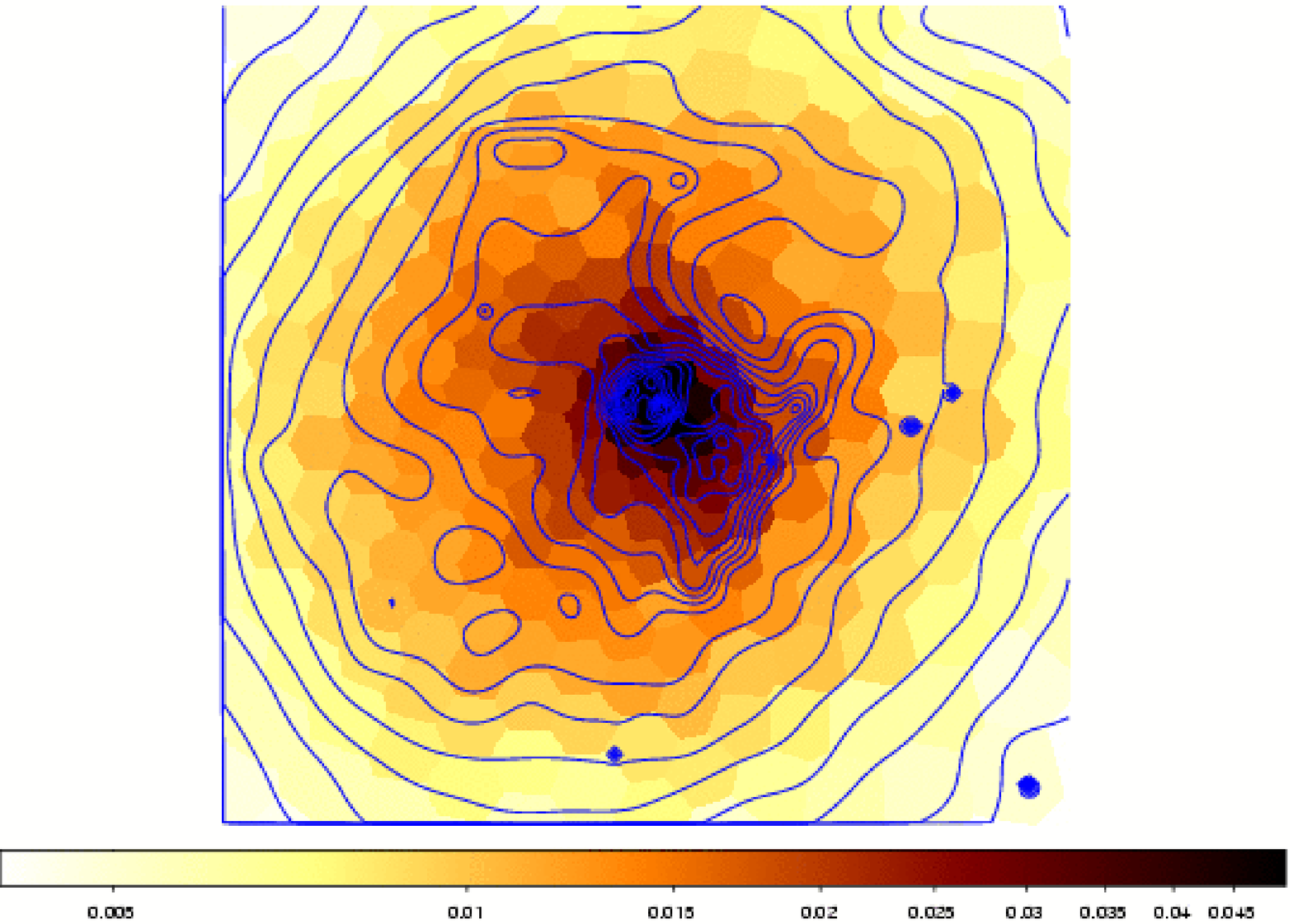}
\includegraphics[width=0.5\textwidth]{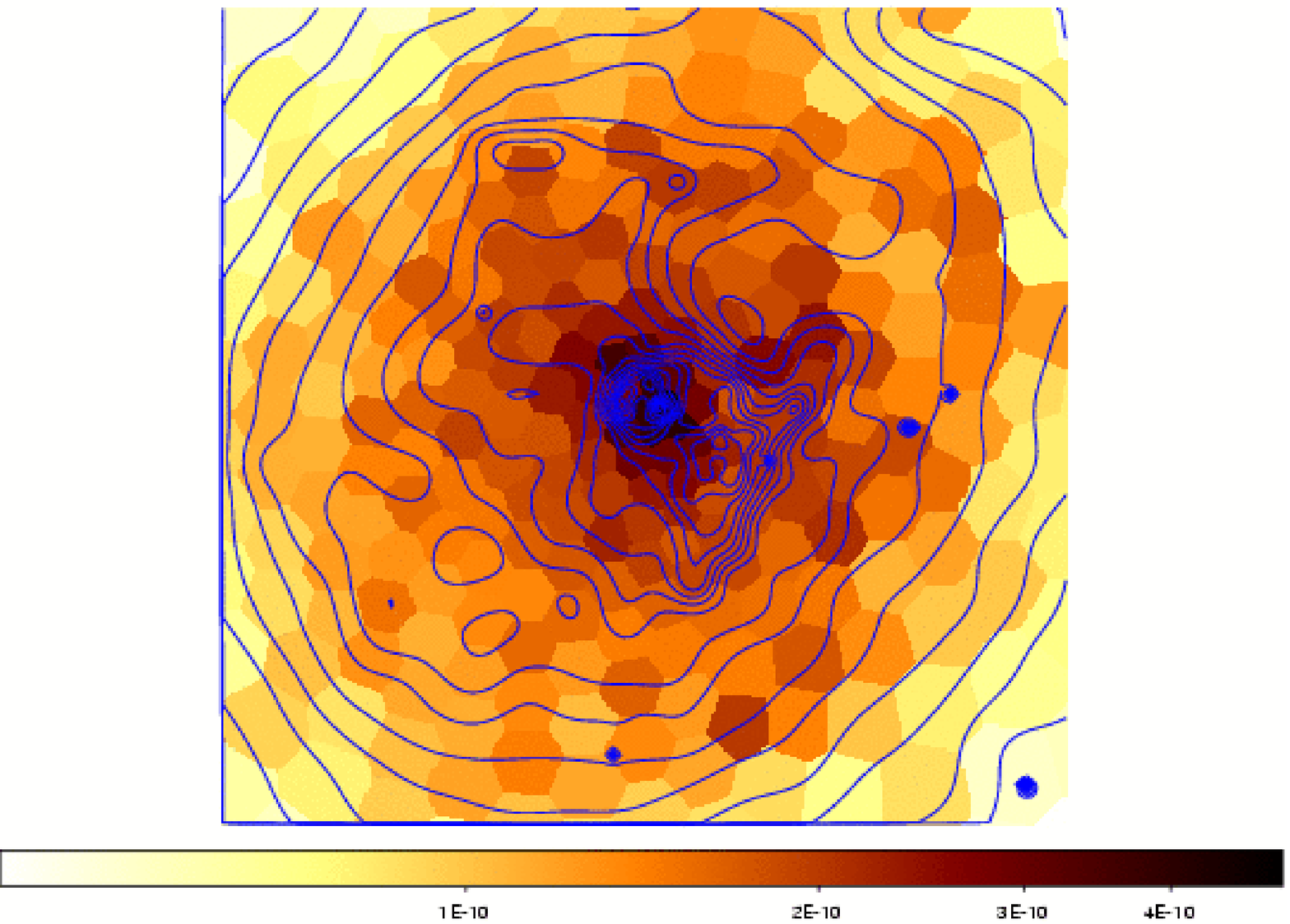}
}
\hbox{
\includegraphics[width=0.5\textwidth]{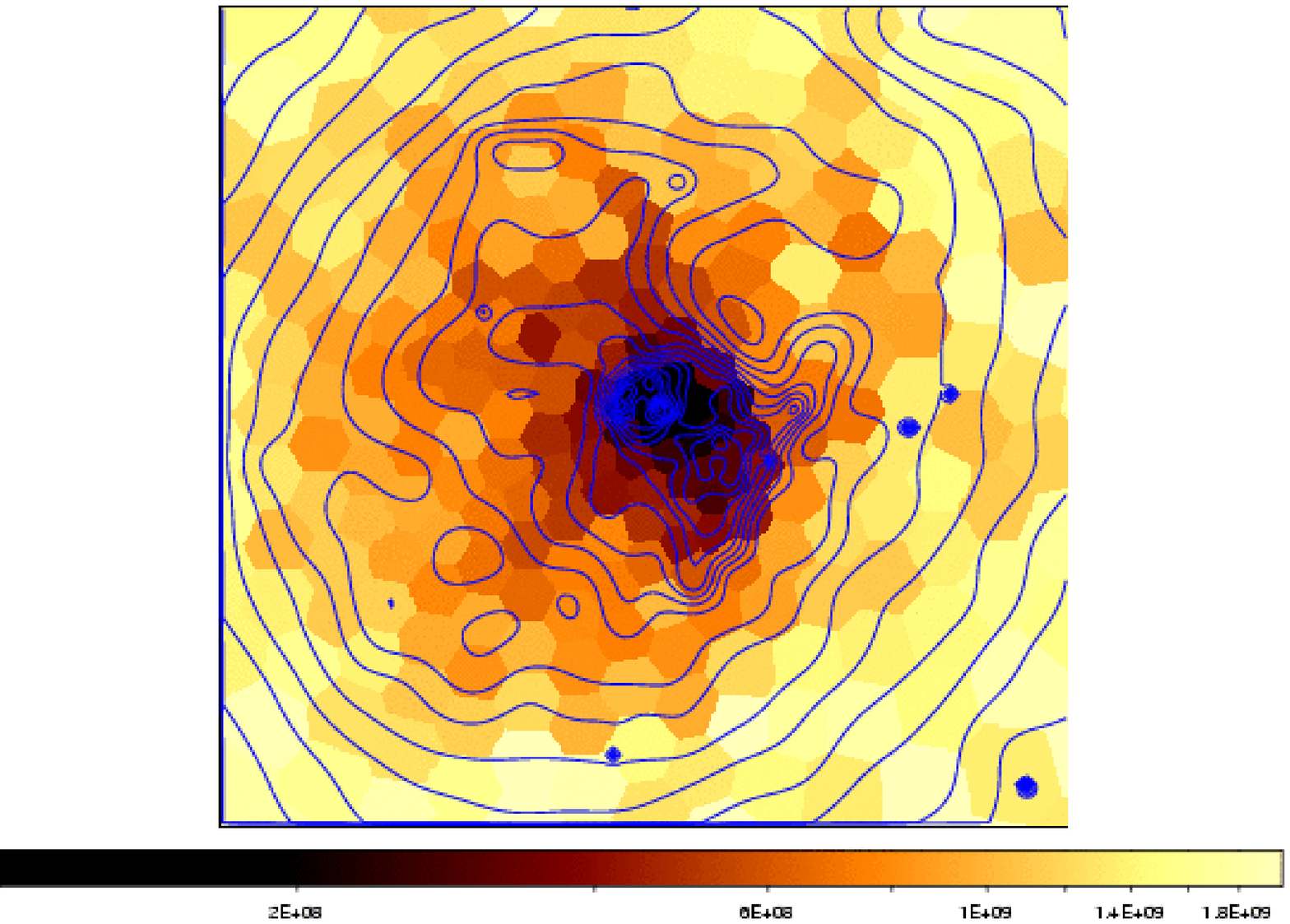}
\includegraphics[width=0.5\textwidth]{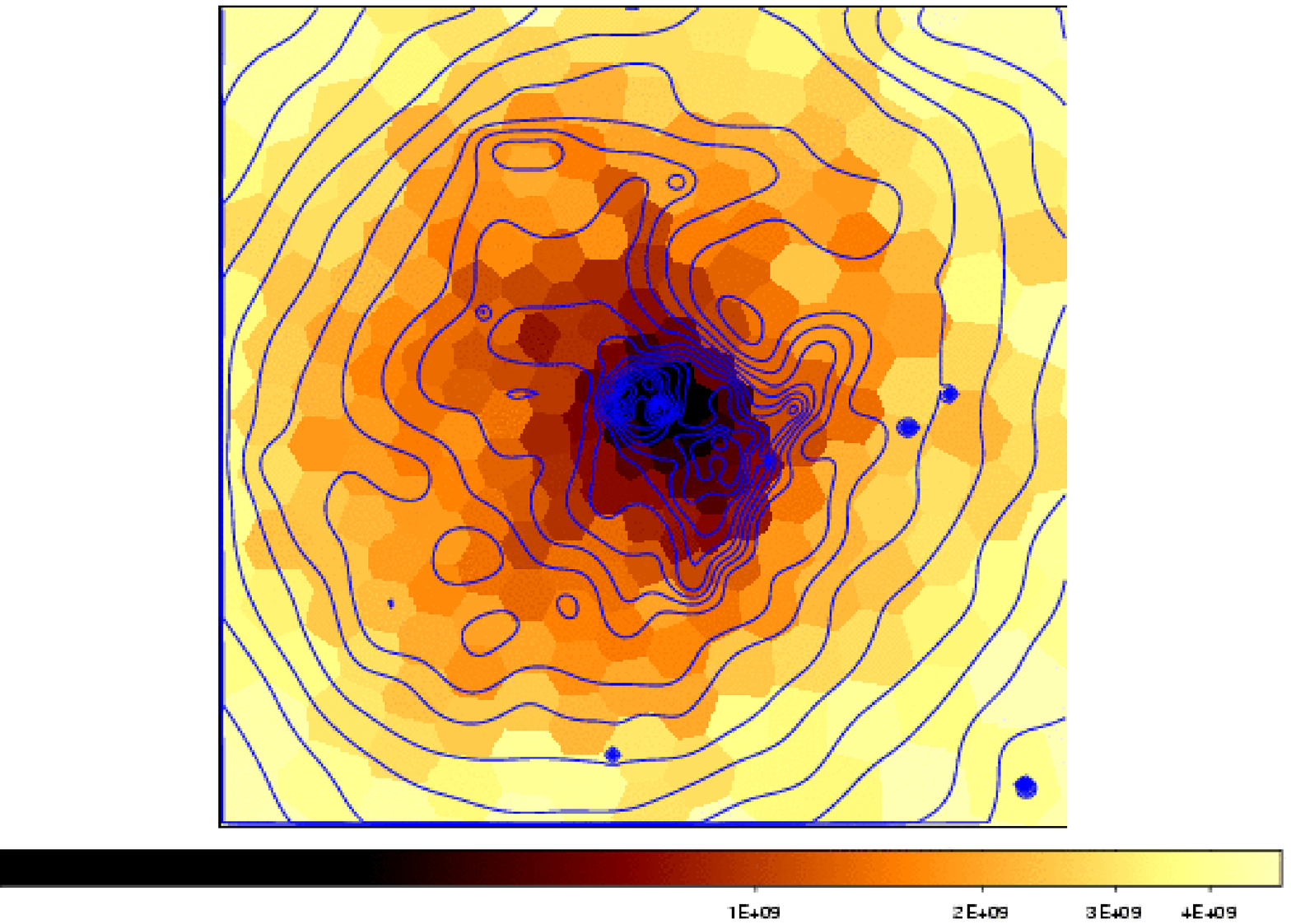}
}
\caption{Maps of density (top-left), pressure (top-right), entropy
index (bottom-left) and cooling time (bottom right).  Color bar units
are cm$^{-2}$ (density), erg\,s$^{-1}$ (pressure), K\,cm$^{-2}$
(entropy index), and years (cooling time).  Contours of the adaptively
smoothed full-band surface brightness are overlaid.}
\vspace{0.5cm}
\label{fig:physical_maps}
\end{figure*}

These maps of density, pressure, entropy index and cooling time are
shown in Fig.~\ref{fig:physical_maps}.  They show the expected trends;
both density and pressure decrease with radius, while entropy and
cooling time increases.  It is particularly interesting to relate
these physical quantities to the SW ridge.  Given the improvement in
the spatial resolution of our maps over those of \citet{choi04}, we
can spatially resolve variations of these physical quantities over
this ridge.  Confirming and strengthening the result of
\citet{choi04}, the pressure map is reasonably symmetric and does not
show any discontinuity coincident with the SW ridge.  Instead, the SW
ridge corresponds to a region of denser, colder gas that maintains a
sharp interface (with a thickness of $<5$\,kpc) with the surrounding
hotter gas.  The gas within the ridge has a cooling time of
500--700\,Myr, compared with cooling times in excess of 1.5\,Gyr in
the ICM it interfaces.

\subsection{Deprojection Analysis}

The 2-d spectral imaging analysis described above does not properly
take into account the full 3-d nature of the cluster.  Each bin will
inevitably include contributions from gas that has a range of
distances from the cluster center.  Hence, any radial gradients of
temperature, density and metalicity will mean that each bin could
therefore include gases with different properties.

To assess the effect of this, we performed a standard deprojection
analysis using the {\tt projct} command within XSPEC.  The observed
cluster emission was divided into 18 concentric, equal width annuli
centered on the core of the cluster PKS2354-35.  Assuming the cluster
to be spherically symmetric (an assumption that clearly breaks down
within the central region), we can ``peel'' spherical shells from the
cluster; the spectrum from a particular shell has the contributions
from more distant shells subtracted, thereby allowing one to model the
spectrum of the gas only in that shell.  We fitted each shell with a
single temperature thermal plasma model ({\tt mekal}) with absorption
fixed at the Galactic value.  The density of the gas in each shell was
calculated from the emission measure (which is directly proportional
to the best-fitting normalization of the {\tt mekal} model) by
assuming that the plasma uniformly filled the shell.  The pressure and
entropy index follows straightforwardly once the temperature and
density are known.

\begin{figure}[t]
\centerline{
\includegraphics[width=0.55\textwidth]{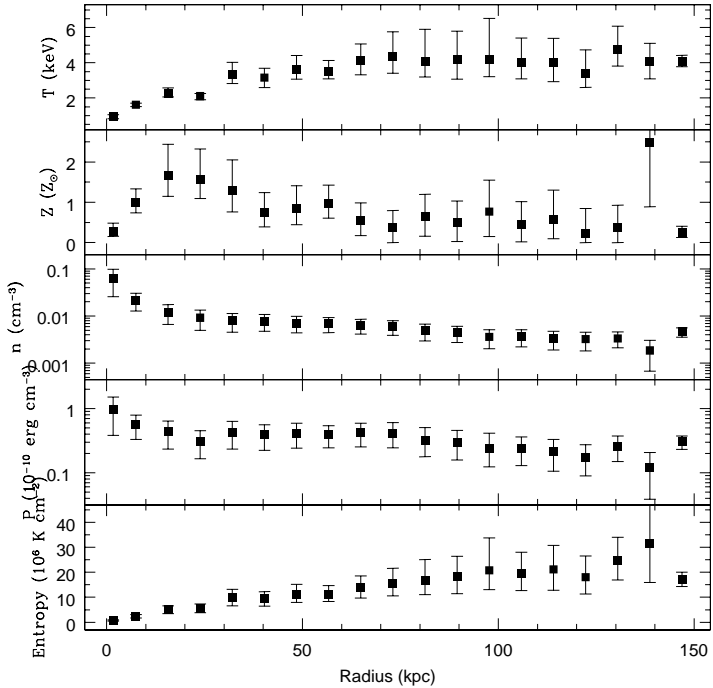}
}
\caption{Single temperature deprojection analysis for A4059.}
\label{fig:deproj}
\end{figure}

\begin{figure}[b]
\centerline{
\includegraphics[width=0.55\textwidth]{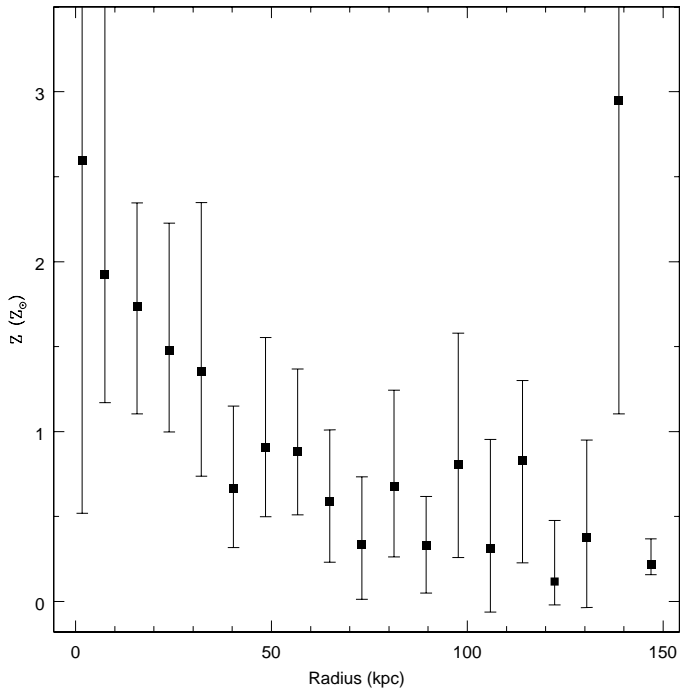}
}
\caption{Metalicity distribution obtained from the two-component
thermal plasma model fitted to the deprojected data for A4059.}
\label{fig:deproj_2t}
\end{figure}

The results of this deprojection are shown in Fig.~\ref{fig:deproj}.
As in \citet{choi04}, the temperature increases with radius, while the
density and pressure decrease. The decrease is is quite rapid along
the inner radii.  Between 10 and 40 kpc, a slight dip in the pressure
is visible, probably due to the presence of the cavities.  The
deprojected metalicity results reflect the behavior seen in the
spectral analysis of the 2-d (projected) data.  The one-component
plasma fit implies a metalicity that is low at the center, increases
to a maximum at $\sim$25\,kpc (corresponding to the metal rich SW
ridge entering the annuli under consideration) and then decreases
again at large radii.  As in the case of the projected spectral
analysis, this central ``hole'' in the metalicity is removed when a
second plasma component (with a different temperature but in pressure
equilibrium and with a common metalicity) is introduced
(Fig.~\ref{fig:deproj_2t}).

\section{Discussion and conclusions}
\label{discussion}

The jet axis of the central radio galaxy PKS~2354--35 is oriented on a
SE to NW line, and this system has approximate reflection symmetry
about a line passing through the AGN perpendicular to the jet axis;
i.e., the jet and counter-jet sides look rather similar.  However,
this system shows striking asymmetries in the direction lateral to the
jet axis. These new data uncover several new aspects of this asymmetry
that were beyond the reach of the previous analyzes of \citet{heinz02}
and \citet{choi04}.

Firstly, both the unsharp mask image (Fig.~\ref{fig:unsharp}) and the
radial surface brightness profile (Fig.~\ref{fig:radial_profile})
reveals a subtle discontinuity in the gradient of the surface
brightness that is approximately semi-circular in form (with
60--80\,kpc radius) and centered on the radio galaxy.  The previous
Chandra observations reported by \citep{heinz02,choi04} had
insufficient signal to detect this feature.  Although it cannot be
rigorously proven from these data, it seems reasonable to interpret
this as a weak shock driven into the previously undisturbed ICM by the
onset of radio galaxy activity; such shocks are seen in all
hydrodynamic simulations of jet/ICM interactions (e.g., Reynolds,
Heinz \& Begelman 2002; Vernaleo \& Reynolds 2006, 2007 and references
therein), and have been found by {\it Chandra} in Hydra A (Nulsen et
al. 2005) and Virgo (Forman et al. 2007).  Assuming that the velocity
of this front through the ICM is greater than the local sound speed
$c_s$ and that the ICM is at rest with respect to the cD galaxy, we
can estimate an upper limit on the time since the initial radio galaxy
activity launched the shock.  If $L$ is the current distance between
this shock front and the AGN, and $t$ is the time since the onset of
radio-galaxy activity launched the shock, we have
   \begin{equation}
   L>\int_{0}^{t} c_s dt, \label{eq:distance}.
   \end{equation}
Using the deprojected temperature profile to determine $c_s(r)$, we
determine that $t<90$\,Myr.  This compares well with estimates for the
source age based on the buoyant rise time of the X-ray cavities
\citep{heinz02}.  We note, however, that we only see this feature on
the NE side of the jet line; there is no similar feature on the SW
side.  This may be due to the simple fact that the ICM emission on the
SW side of the cluster is fainter than the NE side and, hence, the
putative shock is below our detectability limit.  

Secondly, the temperature map shows that the ICM temperature
distribution is strongly asymmetric.  A convex cold front (i.e.,
boundary across which there is a discontinuity in temperature but
continuity in pressure) seems to extend from the central core through
highest surface brightness portions of the SW ridge.  This cooler and
denser material has a comparatively short radiative cooling time, only
500--700\,Myr.  We note that this is rather longer than the
central/ridge cooling times inferred by \citet{choi04}; we consider
our cooling time map to be more robust given that we have the signal
to noise to produce a map in which the gross morphological
structures are resolved [in contrast to \citet{choi04} who could only
examine cooling times within bins that combined core and SW ridge
emission].  One important qualitative difference is that our inferred
cooling times are almost an order of magnitude longer than the
plausible age of the radio source, eliminating the possibility
discussed in \citet{choi04} that the SW ridge is cool due to
compressionally-enhanced radiative cooling.

Thirdly, our new data allow us to map the 2-dimensional distribution
of metals and reveal that it, too, is strongly asymmetric.  High
metalicity gas appears to extend along the length of the SW ridge,
including into the NW spur of the SW ridge that lies beyond the cold
front and may form the wall for the northern cavity.  The result that
the SW ridge possesses enhanced metalicity is robust in both
one-component and two-component thermal plasma fits.  We do find,
however, that the apparent drop in metalicity within the central core
that is inferred from one-component fits is removed when one considers
two-component plasma fits.

Assuming that the jet production process itself is axially symmetric,
all of the evidence points to an initial (pre-interaction) ICM
atmosphere that was strongly asymmetric.  The asymmetries in this
system have previously been interpreted in terms of a bulk flow within
the ICM (flowing in the plane of the sky from the SW to the NE) prior
to the onset of powerful radio galaxy activity.  However, it is very
hard to understand the metalicity asymmetry in terms of such a bulk
flow.  Instead, it seems likely that an anomalously dense, cool,
metal-rich region of gas was situated to the SW of the center of the
cD galaxy at the time that the AGN initiated its jet activity.  This
anomalous region would have been weakly shocked and possibly pushed
higher into the ICM atmosphere by the laterally expanding radio galaxy
cocoon, thereby forming the SW ridge.  If this scenario is correct, it
seems that some fraction of this metal rich gas has been entrained
into the walls of the X-ray cavities, especially in the NW direction.

At this point, it is useful to estimate the mass of gas within the SW
ridge and the gravitational potential energy of this structure.  The
brightest region of the (projected) SW ridge emission can be
encompassed by an ellipse with major and minor axes of 15\,arcsec and
8\,arcsec respectively.  Assuming that this is a oblate spheroid in
projection, we estimate the emitting volume to be $V=2\times
10^{68}\,{\rm cm}^3$.  A single-component thermal plasma fit provides an
adequate description of the data and implies a total emission measure
of
\begin{equation}
EM\equiv \int_Vn_e^2\,dV\approx 1\times 10^{65}\,{\rm cm}^{-3}.
\end{equation}
Assuming that twice cosmic abundance plasma uniformly fills this
region, we conclude that the total mass of ICM within this ridge is
$M\sim 5\times 10^9\,{\rm M}_\odot$.  We can also estimate the energy
required to lift this matter from a location close to the center of
the cD galaxy to its current position in the gravitational potential.
Assuming that the undisturbed ICM is approximately isothermal with
sound speed $c_s$ and is in a hydrostatic configuration with density
profile $\rho({\bf r})$, integration of the equation of hydrostatic
equilibrium gives the gravitational potential,
\begin{equation}
\Phi=\frac{c_s^2}{\gamma}\ln\left(\frac{\rho}{\rho_0}\right),
\end{equation}
where $\gamma= 5/3$ is the ratio of specific heat capacities and
$\rho_0$ is some fiducial density that plays the role of the
(uninteresting) constant of integration.  The change in gravitational
potential energy when mass $M$ is lifted within the atmosphere is
given by
\begin{equation}
\Delta E=M\Delta\Phi=\frac{Mc_s^2}{\gamma}\ln\left(\frac{\rho_1}{\rho_2}\right),
\end{equation}
where $\rho_1$ and $\rho_2$ are the densities of the undisturbed ICM
atmosphere at the start point and end point, respectively, of the
lifting process.  Using values of $c_s\approx 700\,{\rm km}\,{\rm
s}^{-1}$ (corresponding to to the $kT\approx 2$\,keV plasma that
characterizes the central regions of the ICM), and $\rho_1/\rho_2\sim
5$ (from the deprojection results), we obtain $\Delta E\sim 5\times
10^{58}\,{\rm erg}$.  To place this energy into context,
\citet{heinz02} estimated that the energy required to inflate the
observed X-ray cavities exceeded $8\times 10^{59}\,{\rm erg}$, almost
20 times greater than the energy required to lift the cool gas in the
SW ridge.

While it seems clear that the ICM atmosphere was asymmetric prior to
the (most recent) onset of radio galaxy activity, the cause of that
asymmetry is unclear.  One possibility is that the asymmetry was
caused by the merger/assimilation of A4059 with a small galaxy cluster
or group.  The SW ridge could then be identified with the remains of
the ICM core of the smaller subcluster (albeit with a morphology that
has been subsequently shaped by interaction with the jet-blown cocoon
of PKS~2354--35).  Indeed, Heinz et al. (2003) have shown that some
fraction of ICM cold fronts are likely to be associated with
subcluster merger events, and that the hydrodynamics of such an
interaction can cause high metalicity gas to be dredged up from the
center of the merging core and carried up to the cold front.  This
would qualitatively provide a natural explanation for the high
metalicity of the SW ridge.  The gross X-ray morphology of A4059 is
also reminiscent of the ``X-ray comet tail'' found in the Abell 2670
cluster (Fujita, Sarazin \& Sivakoff 2006).  The mass of the gas in
the ``comet tail'' is similar to the SW ridge in A4059, and it also
appears cooler than the ambient ICM.  In the case of A2670 the
association of the ``comet tail'' with a large elliptical galaxy, and
the high peculiar motion of the cD galaxy, makes it very plausible
that we are witnessing the merger of a small cluster or group with
A2670 and that the comet-tail is the distorted/stripped remnants of
the subcluster's ICM core.

The subcluster merger scenario faces two potential problems.  Firstly,
unlike the case of A2670, there is no obvious association of the SW
ridge with any galaxy.  Since the galaxies associated with the merging
subcluster may have left behind their ICM core (such as is seen in the
Bullet Cluster 1E~0657--56; Markevitch et al. 2002), this fact by
itself does not rule out the merging subcluster hypothesis; it does
predict, however, that a detailed mapping of the galaxy dynamics in
this cluster should reveal a kinematically distinct subcluster.
Secondly, the metalicity of the SW ridge is $1.5-2\,{\rm Z}_\odot$, an
extreme value if this is to be identified with the ICM core of a
merging subcluster (Baumgartner et al. 2005; Snowden et al. 2007).
Hence, while the merging subcluster hypothesis cannot yet be ruled
out, we are led to explore other scenarios.

The very high metalicity of the SW ridge, in particular, pushes us to
explore scenarios in which the gas originates directly from the metal
enriched ISM of an actively star-forming galaxy.  In order to deposit
$5\times 10^9\,{\rm M}_\odot$ of metal rich gas into a localized
region of the cluster, the donor galaxy must have been massive
(comparable to or greater than an $L^*$ galaxy) and the stripping
mechanism must have been efficient and rapid.  One possibility is that
a large gas-rich late-type galaxy plunged through the core of A4059
for the first time and underwent an intense compressionally-driven
starburst (Reverte et al. 2007).  The combination of the
starburst-driven superwind and ram-pressure stripping by the ICM could
have allowed the galaxy to dump the required amount of metal-rich gas
on a timescale short compared with the time taken to traverse the
cluster core.  Essentially, the star-bursting galaxy would leave a
``debris'' trail behind it comprising of metal rich, relatively cool,
possibly multiphase gas.  We hypothesize that this debris trail is the
high-metalicity inhomogeneity which, once impacted by the radio-galaxy
activity, produces the SW ridge feature in A4059.

\begin{figure*}[t]
\centerline{
\includegraphics[width=0.9\textwidth]{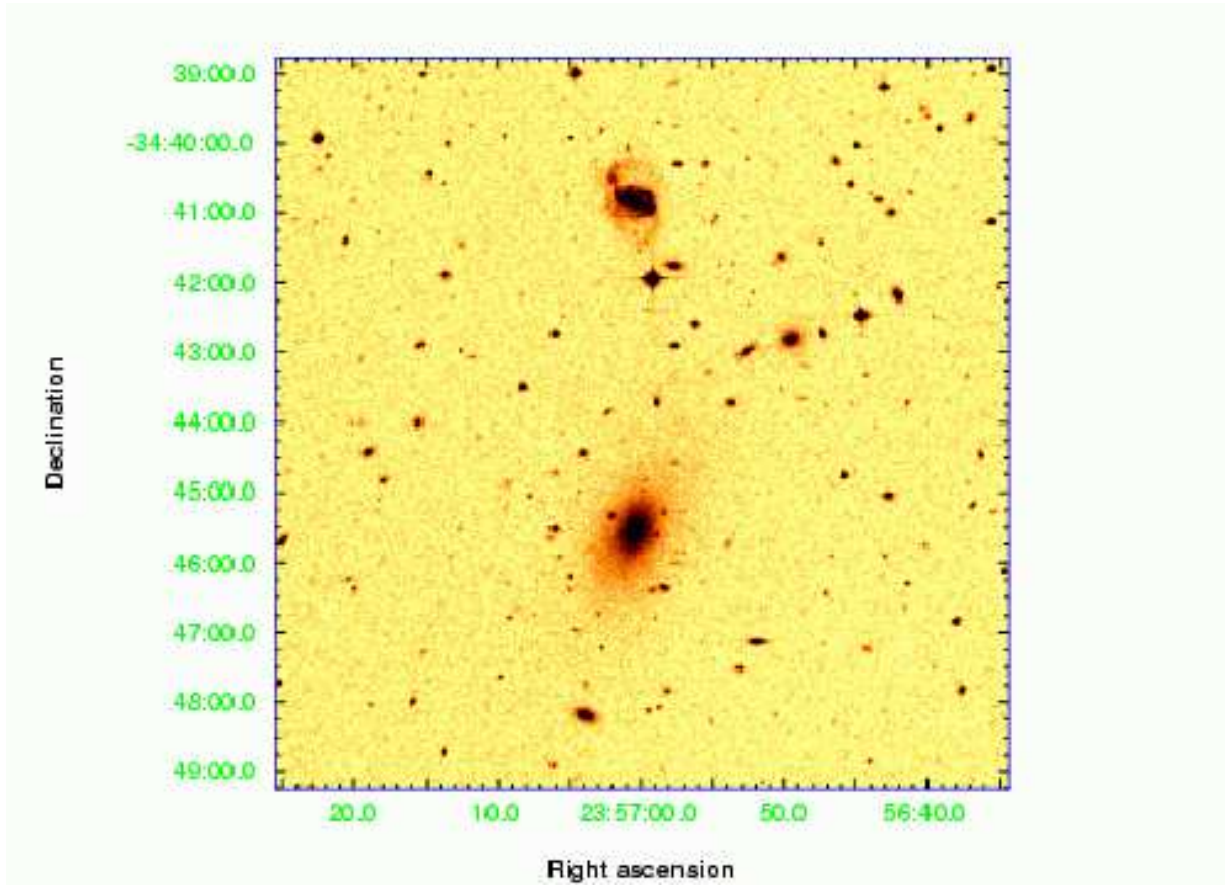}
}
\caption{Digitized Sky Survey image of a $10\times 10\,{\rm
arcmin}^2$ field in the central regions of the A~4059 cluster.  The
cD galaxy ESO~349--G010 hosts the radio-galaxy PKS~2354--35 and
resides at the center of the ICM atmosphere.  The bright spiral galaxy
ESO~349--G009 can be clearly seen to the north of the cD galaxy.}
\vspace{0.5cm}
\label{fig:dss_field}
\end{figure*}

If the projected (i.e., plane of the sky) velocity of the galaxy
during its passage through the cluster core is $v=1000v_3\,{\rm
km}\,{\rm s}^{-1}$, the metal rich ISM would have to be stripped from
the galaxy at a rate exceeding $200v_3\,{\rm M}_\odot {\rm yr}^{-1}$
in order produce a distinct region consisting of $5\times 10^9\,{\rm
M}_\odot$ of metal-rich gas that is only $\sim 25$\,kpc in projected
extent.  Once stripped away from the galaxy, this denser gas will
follow a decaying orbit within the (dark matter dominated)
gravitational potential due to the action of gas drag.  Concurrently,
hydrodynamic instabilities will tend to mix this stripped gas with the
ambient ICM.  Detailed theoretical modeling beyond the scope of this
paper is required in order to estimate the time over which the
starburst-delivered gas will remain a distinct inhomogeneity within
the ICM core.  For now, we estimate an upper limit on the time since
this starburst/stripping event occurred by equating it with the
orbital time of the gas within the potential, $\tau\sim
2\pi\sqrt{r/g}$, where $g\equiv -\nabla\Phi$ is the acceleration due
to gravity in the cluster potential.  This timescale is within a
factor of order unity the same as the time taken for the cooler gas to
fall radially into the center of the potential.  Assuming that the
underlying ambient ICM density field is approximately spherically
symmetric $\rho=\rho(r)$, we have
\begin{equation}
g=-\frac{c_s^2}{\gamma r}\frac{r\,d\rho}{\rho\,dr},
\end{equation}
and we see that
\begin{equation}
\tau\sim 2\pi\sqrt{\frac{r^2\gamma}{c_s^2\,(-r\,d\rho/\rho\,dr)}}.
\end{equation}
Putting $r=25$\,kpc, and noting that $-r\,d\rho/\rho\,dr\sim 1-2$, we
estimate that the starburst/stripping event occurred at most $\tau\sim
200-300\,{\rm Myr}$ ago.

An obvious and important prediction of this starburst-stripping
hypothesis is that the responsible galaxy should still be in the
vicinity and should still display post-starburst signatures.  Using
the above estimate of the upper limit on the time since the
starburst/stripping event, the post-starburst galaxy should still be
within a projected distance of $300v_3\,{\rm kpc}$ of the cluster
core.  This corresponds to an angular distance of
$285v_3$\,arcsec.  Of course, deceleration of the galaxy as it
climbs out of the dark matter potential would reduce this distance.

A systematic study of the optical spectrum of the member galaxies of
A4059 is required in order to find the post-starburst galaxy predicted
by this hypothesis.  One interesting candidate, however, is the bright
spiral galaxy ESO~349--G009 situated $280^{\prime\prime}$ directly to
the north of the cD galaxy ESO~349--G010 (see
Fig.~\ref{fig:dss_field}).  The radial velocity difference between
ESO~349--G009 and ESO~349--G010/PKS2354--35 is almost $\Delta
v=2100\,{\rm km}\,{\rm s}^{-1}$ ($v=12612\,{\rm km}\,{\rm s}^{-1}$ for
ESO~349--G009 compared with $v=14705\,{\rm km}\,{\rm s}^{-1}$ for
ESO~349--G010/PKS2354--35; de Vaucouleurs et al., 1991) implying that
this late-type galaxy is either simply a foreground object or a
high-velocity galaxy close to the A4059 cluster.  From the absolute
2MASS K-band magnitude of $M_K=-25.82$, we can assume a K-band
stellar-mass to light ratio of 0.7 (Bell et al. 2003) and a solar
K-band absolute magnitude of $3.27$ (Holmberg et al. 2006) in order to
derive an estimate for the stellar mass of this galaxy, $M_*\approx
3\times 10^{11}\,{\rm M}_\odot$.  Thus, we see that the SW ridge in
the ICM core could have been formed via the ejection of gas totaling a
few percent of the stellar mass.

\begin{figure}[b]
\centerline{
\includegraphics[width=0.5\textwidth]{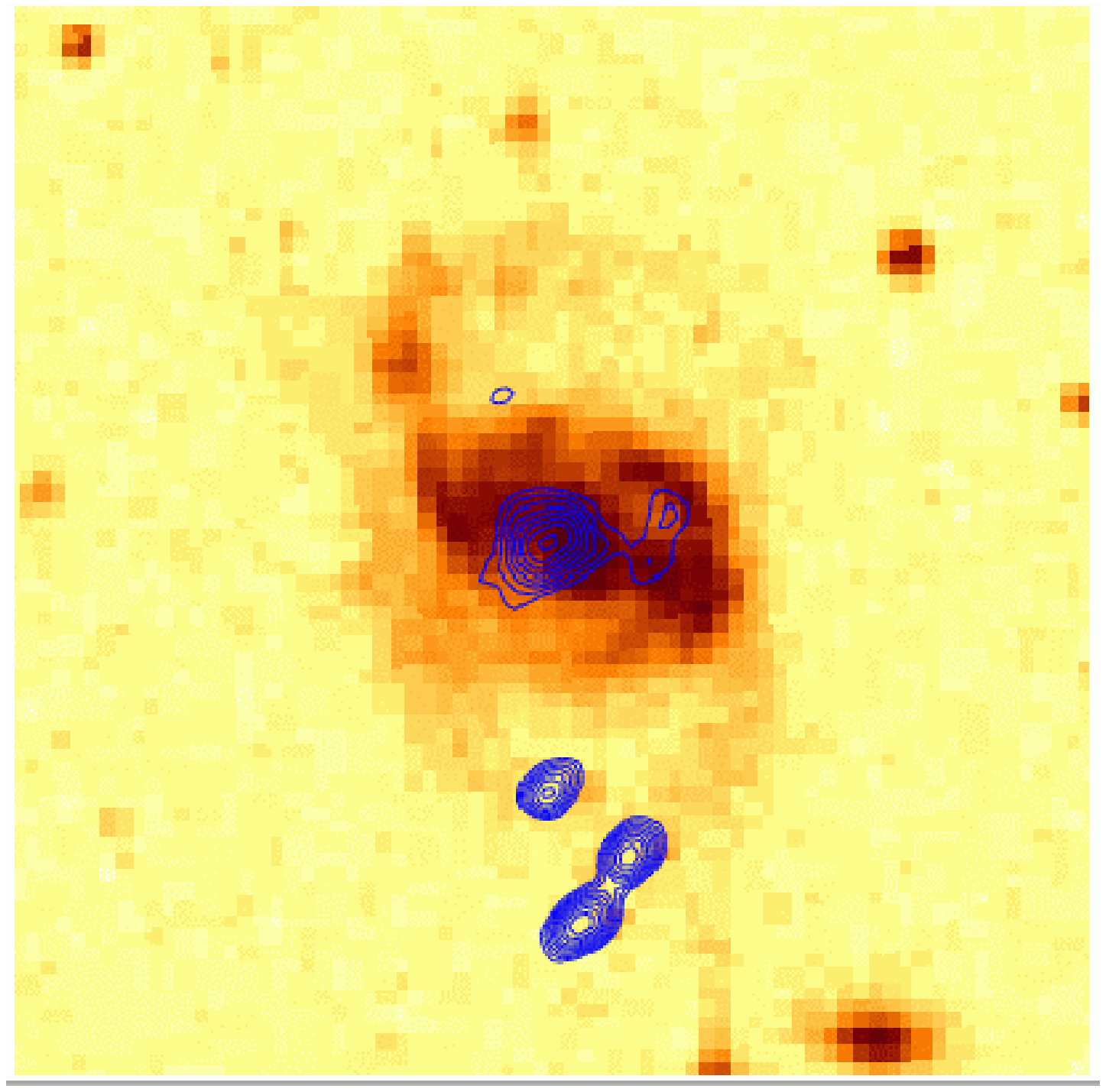}
}
\caption{Digitized Sky Survey image of a $2\times 2\,{\rm arcmin}^{2}$
field centered on ESO~349--G009.  Overlaid are contours of the
full-band (0.5--10\,keV) X-ray intensity.  See text and Sun et
al. (2007) for a discussion of this emission.}
\vspace{0.5cm}
\label{fig:ngal}
\end{figure}

What makes ESO~349--G009 a particular interesting candidate is the
presence of X-ray and near ultraviolet (NUV) evidence for on-going
vigorous star formation.  Sun et al. (2007) present an analysis of the
X-ray emission from ESO~349--G009 using the same {\it Chandra} data as
reported upon here.  They find a luminous (spatially resolved) X-ray
corona associated with the galactic bulge in addition to three point
sources associated with the spiral arms (Fig.~\ref{fig:ngal}).  They
also note that, despite the absence of an AGN (i.e., an unresolved
source at the location of the galactic nucleus) in this galaxy, the
X-ray emission has a hard component with $L_{2-10\,keV}=0.8-2.7\times
10^{41}\,{\rm erg}\,{\rm s}^{-1}$.  Associating the hard X-rays with
High Mass X-ray Binaries, they use the correlation of Grimm, Gilfanov
\& Sunyaev (2003) to estimate a star formation rate of $>12\,{\rm
M}_\odot{\rm yr}^{-1}$.  This conclusion is supported by the fact that
ESO~349--G009 is luminous in the NUV band (see Fig.~26 of Sun et
al. 2007).  Within the context of our starburst-stripping hypothesis,
one can postulate a scenario in which ESO~349--G009 executed a high
velocity plunge through the core of A4059 and suffered an intense
ICM-pressure induced starburst which is still in the process of dying
away.  High-quality optical spectroscopy of ESO~349--G009 combined
with stellar-synthesis models will allow its recent star formation
history to be constrained and will provide a powerful check of this
scenario.  In particular, an intense starburst 100--300\,Myr ago will
be revealed through strong Balmer line absorption features resulting
from the large numbers of A stars.

It must be noted, however, that even just the current line-of-sight
velocity difference between ESO~349--G009 and
ESO~349--G010/PKS2354--35 likely requires that ESO~349--G009 is
marginally unbound to A~4059.  Thus, if this galaxy is indeed the
origin of the high metallicity plume, this system is a rare example of
an unbound massive galaxy falling through the core of a rich cluster.   

The temporal coincidence between the onset of powerful radio-galaxy
activity and the hypothesized plunge/stripping of a large starburst
galaxy suggests a connection.  There are two principal possibilities.
The close passage of a high velocity massive galaxy would
tidally-shock any cold gas within the cD galaxy, driving an inflow and
possibly triggering AGN activity.  Alternatively, the gas stripped
from the starburst galaxy is very likely to be multiphase (with
regions of cold, atomic gas entrained by the hot X-ray emitting gas).
Once stripped from the passing galaxy, the cold and dense atomic gas
would preferentially fall (or, more precisely, sediment) into the cD
galaxy and, again, possibly fuel the AGN activity.  Either way, this
presents a very different model of ``radio-mode'' AGN feedback to that
normally envisaged; in this case, we are witnessing AGN feedback
triggered directly by the close passage of a gas rich galaxy rather
than radiative cooling of the host ICM.

\acknowledgments We thank Steve Allen, Andrew Fabian, Stacy McGaugh,
Richard Mushotzky David Rupke, and Kimberly Weaver for useful
discussions. We also thank the anonymous referee for insightful
comments that significantly improved this paper.  Finally, we are
grateful to S.~Diehl and T.S.~Statler, whose WVT binning algorithm we
utilized.  This was a generalization of Cappellari \& Copin's (2003)
Voronoi binning algorithm.  This work was supported by Smithsonian
Astrophysical Observatory grant GO5-6128X.


\bibliographystyle{apj}

\end{document}